\documentstyle[leqno,12pt]{article}

\if@twoside
\else
\fi
\advance\textheight by \topskip
\leftmargin\leftmargini
\labelwidth\leftmargini\advance\labelwidth-\labelsep

\catcode`\@=11

\def\Let@{\relax\iffalse{\fi\let\\=\cr\iffalse}\fi}
\def\vspace@{\def\vspace##1{\crcr\noalign{\vskip##1\relax}}}
\def\multilimits@{\bgroup\vspace@\Let@
 \baselineskip\fontdimen10 \scriptfont\tw@
 \advance\baselineskip\fontdimen12 \scriptfont\tw@
 \lineskip\thr@@\fontdimen8 \scriptfont\thr@@
 \lineskiplimit\lineskip
 \vbox\bgroup\ialign\bgroup\hfil$\m@th\scriptstyle{##}$\hfil\crcr}
\def\Sb{_\multilimits@}
\def\endSb{\crcr\egroup\egroup\egroup}
\def\Sp{^\multilimits@}

\long\def\QQQ#1#2{}
\def\QTP#1{}
\long\def\QQA#1#2{}

\def\EXPAND#1[#2]#3{}
\def\NOEXPAND#1[#2]#3{}

\def\LaTeXparent#1{}

\QQQ{Language}{
American English
}

\begin{document}

$ $

\vskip 3cm

\begin{center}
\LARGE Hamilton-Jacobi formulation for singular systems with second order
Lagrangians
\end{center}

\,

\begin{center}
B. M. Pimentel\footnote{%
Partially supported by CNPq} and R. G. Teixeira\footnote{%
Supported by CAPES}\\
\end{center}

\,

\begin{center}
Instituto de F\'{\i}sica Te\'orica\\Universidade Estadual Paulista\\Rua
Pamplona 145\\01405-900 - S\~ao Paulo, S.P.\\Brazil\\
\end{center}

\newpage
$ $

\vskip 5cm

\begin{center}
\begin{minipage}{14.5cm}
\centerline{\bf Abstract}
\,
Recently the Hamilton-Jacobi formulation for first order constrained systems
has been de\-ve\-lo\-ped. In such formalism the equations of motion are
written as total differential equations in many variables.
We generalize the Hamilton-Jacobi formulation for singular systems with
se\-cond order Lagrangians and apply this new formulation to Podolsky
electrodynamics, comparing with the results obtained through Dirac's
method.
\end{minipage}
\end{center}

\vskip 1.5cm



\newpage\

\section{Introduction}

Systems with higher order Lagrangians have been studied with increasing
interest because they appear in many relevant physical problems. As examples
we have the consistent regularization of ultraviolet divergences in
gauge-invariant supersymmetric theories \cite{11} or effective Lagrangians
in gauge theories \cite{12}. Besides this, the fact that gauge theories have
singular Lagrangians is in itself a motivation to the study of the formalism
for second order singular Lagrangians.

The Lagrangian formulation for constrained systems can be found in
references \cite{6} and \cite{7} while the Hamiltonian formulation of
singular systems is usually made through a formalism developed by Dirac \cite
{3,4,5}. In this formalism the constraints caused by the Hessian matrix
singularity are added to the canonical Hamiltonian and then the consistency
conditions are worked out, being possible to eliminate some degrees of
freedom of the system. Dirac also showed that the gauge freedom is caused by
the presence of first class constraints.

The study of new formalisms for singular systems may provide new tools to
investigate these systems. In classical dynamics, different formalisms
(Lagrangian, Hamiltonian, Hamilton-Jacobi) provide different approaches to
the problems, each formalism having advantages and disadvantages in the
study of some features of the systems and being equivalent among themselves.
In the same way, different formalisms provide different views of the
features of singular systems, which justify the interest in their study.

Here we generalize the Hamilton-Jacobi formalism that was recently developed
\cite{1,2} to include singular second order Lagrangians. We start in Sect. 2
with an overview of the features of singular systems with second order
Lagrangians and of the Dirac's Hamiltonian formalism for them. In Sect. 3 we
develop the Hamilton-Jacobi formulation for a general second order system
and apply this formalism to the case of a singular second order system in
Sect. 4. An example is solved using both Dirac's and Hamilton-Jacobi
formalism in Sect. 5, while Sect. 6 is devoted to the conclusions.

\section{Singular systems with second order Lagran\-gians}

The treatment for theories with higher order derivatives has been first
developed by Ostrogradski \cite{8} and allows to write the Euler-Lagrange
equations, introduce conjugated momenta and develop a Hamiltonian formalism
for such systems. Here we center our attention on the case of a system
described by a Lagrangian containing time derivatives of the coordinates up
to second order (a second order Lagrangian).

In such a theory the configuration space is described by the $n$ generalized
coordinates $q_i$ and its first and second derivatives with respect to the
time parameter $t$ (we consider a discrete system to simplify the
calculations; the generalization to continuous systems is straightforward
and will be done in Sect. 5).

The Euler-Lagrange equations, which are obtained from the action integral
\begin{equation}  \label{Eq.1}
S=\int L\left( q,\stackrel{.}{q},\stackrel{..}{q},t\right) dt
\end{equation}
using the Hamilton's principle, are given by:
\begin{equation}  \label{Eq.2}
\frac{\partial L}{\partial q_i}-\frac d{dt}\left( \frac{\partial L}{\partial
\stackrel{.}{q}_i}\right) +\frac{d^2}{dt^2}\left( \frac{\partial L}{\partial
\stackrel{..}{q}_i}\right) =0
\end{equation}

We construct the phase space by introducing the generalized momenta
\begin{equation}  \label{Eq.3}
p_i=\frac{\partial L}{\partial \stackrel{.}{q}_i}-\frac d{dt}\left( \frac{%
\partial L}{\partial \stackrel{..}{q}_i}\right)
\end{equation}
\begin{equation}  \label{Eq.4}
\pi _i=\frac{\partial L}{\partial \stackrel{..}{q}_i}
\end{equation}
(conjugated respectively to $q_i$ and $\stackrel{.}{q}_i$) and writing the
accelerations $\stackrel{..}{q}_i$ as functions of the coordinates $q$,
velocities $\stackrel{.}{q}$ and of the momenta $p$ and $\pi $ ($\stackrel{..%
}{q}_i=f_i\left( q_i,\stackrel{.}{q_i},p_i,\pi _i\right) $). The phase space
will then be spanned by the canonical variables $\left( q_i,p_i\right)
,\left( \stackrel{\_}{q}_i,\pi _i\right) $ where $\stackrel{\_}{q}_i=%
\stackrel{.}{q}_i$.

By introducing the canonical Hamiltonian defined as
\begin{equation}  \label{Eq.4a}
H_C=p_i\stackrel{\_}{q}_i+\pi _i\left. \stackrel{..}{q}_i\right| _{\stackrel{%
..}{q}_i=f_i}-\left. L\right| _{\stackrel{..}{q}_i=f_i}
\end{equation}
we can write the equations of motion of any function $g$ of the canonical
variables as:
\begin{equation}  \label{Eq.4b}
\stackrel{.}{g}=\left\{ g,H_C\right\}
\end{equation}

But this procedure is only possible if the determinant of the Hessian matrix
\begin{equation}
H_{ij}=\frac{\partial ^2L}{\partial \stackrel{..}{q}_i\partial \stackrel{..}{%
q}_j}  \label{Eq.5}
\end{equation}
does not vanish, otherwise it will not be possible to express all the
accelerations $\stackrel{..}{q}_i$ as functions of the canonical variables
and there will be relations as
\begin{equation}
\Phi _\alpha \left( q_i,p_i;\stackrel{\_}{q}_i,\pi _i\right) =0;\;\alpha
=1,...,m<2(n-1)  \label{Eq.6}
\end{equation}
connecting the momenta variables. As a consequence we will not be able to
treat the canonical variables as an independent set and we have to retreat
to formalisms specially developed to deal with the dependence among the
canonical variables, i.e. a formalism for constrained systems.

The usual treatment of singular systems was developed by Dirac \cite{3,4,5}
who, in order to deal with this problem, introduced a generalized
Hamiltonian formalism (details can be found in references \cite{6}, \cite{7}
and \cite{9}). Such formalism can also be applied to the case of second
order Lagrangians as can be seen in references \cite{9} and \cite{10}.
Dirac's formalism consist in considering constraints given by equation (\ref
{Eq.6}) as {\it weak} equations, called primary constraints, and represented
as:
\begin{equation}
\Phi _\alpha \left( q_i,p_i;\stackrel{\_}{q}_i,\pi _i\right) \approx 0
\label{Eq.7}
\end{equation}

By weak equations we mean those that can't be used until all Poisson
brackets have been calculated.

We may add to the canonical Hamiltonian $H_C$ any linear combination of the
primary cons\-traints and define a new Hamiltonian, called total
Hamiltonian, given by
\begin{equation}
H_T=H_C+u_\alpha \Phi _\alpha ,  \label{Eq.8}
\end{equation}
where $u_\alpha $ are arbitrary coefficients. Physically $H_C$ and $H_T$ are
equivalent and we cannot distinguish between them. The equation of motion
for any function $f\left( q_i,\stackrel{\_}{q}_i,p_i,\pi _i\right) $ is
given in terms of $H_T$ as:
\begin{equation}
\stackrel{.}{f}\approx \left\{ f,H_T\right\} =\left\{ f,H_C\right\}
+u_\alpha \left\{ f,\Phi _\alpha \right\}  \label{Eq.9}
\end{equation}

The constraints will produce consistency conditions because they must be
valid at any time and consequently their time derivative must be weakly
zero. The consistency conditions are given by:
\begin{equation}  \label{Eq.13}
\stackrel{.}{\Phi }_\alpha \approx \left\{ \Phi _\beta ,H_T\right\} =\left\{
\Phi _\beta ,H_C\right\} +u_\alpha \left\{ \Phi _\beta ,\Phi _\alpha
\right\} \approx 0;\;\alpha ,\beta =1,...,m
\end{equation}

These conditions may be either identically satisfied (when we use the
primary constraints), determine some of the arbitrary coefficients $u$, or
generate new constraints that will be called secondary constraints. The
constraints that have null Poisson brackets with all other constraints are
called first class constraints otherwise they are called second class ones.
This classification is completely independent of the division in primary and
secondary constraints. The extended Hamiltonian is defined as
\begin{equation}  \label{Eq.13b}
H_E=H_C+V_\lambda \Psi _\lambda
\end{equation}
were the $\Psi _\lambda $ include all first class constraints. $V_\lambda $
are arbitrary coefficients and we use (\ref{Eq.13b}) instead of the total
Hamiltonian (\ref{Eq.8}) in the equations of motion.

\section{Hamilton-Jacobi formalism for second order La\-gran\-gians}

Recently a new formalism for singular first order systems was developed by
G\"uler \cite{1,2} who obtained a set of Hamilton-Jacobi partial
differential equations for such systems using Carath\'eo\-do\-ry's
equivalent Lagrangians method and wrote the equations of motion as total
differential equations.

In this section we will use Carath\'eodory's method to develop the
Hamilton-Jacobi for\-ma\-lism to a general second order Lagrangian. This
formalism can be applied to any second order Lagrangian and is not limited
to singular ones. The singular case will be considered in the next section.

Carath\'{e}odory's equivalent Lagrangians method to second order Lagrangians
says that, given a Lagrangian $L(q_i,\stackrel{.}{q}_i,\stackrel{..}{q}_i,t)$%
, we can obtain a completely equivalent one by:
\begin{equation}
L^{\prime }=L\left( q_i,\stackrel{.}{q}_i,\stackrel{..}{q}_i,t\right) -\frac{%
dS\left( q_i,\stackrel{.}{q}_i,t\right) }{dt}  \label{Eq.14}
\end{equation}

These Lagrangians are equivalent because the action integral given by them
have simultaneous extremes. So we can choose the function $S(q_i,\stackrel{.%
}{q}_i,t)$ in such a way that $L^{\prime }$ becomes an extreme and then we
reduce the variational problem of finding extreme for the Lagrangian $L$ to
a problem of differential calculus. To do this we must find a set of
functions $\varphi _i(q_i,\stackrel{.}{q}_i,t)$ , $\beta _i(q_i,t)$ and $%
S(q_i,\stackrel{.}{q}_i,t)$ such that
\begin{equation}
L^{\prime }\left( q_i,\beta _i,\varphi _i,t\right) =0  \label{Eq.15}
\end{equation}
and for all neighborhood of $\stackrel{.}{q}_i=\beta _i(q_i,t)$ and $%
\stackrel{..}{q}_i=\varphi _i(q_i,\stackrel{.}{q}_i,t)$:
\begin{equation}
L^{\prime }\left( q_i,\stackrel{.}{q}_i,\stackrel{..}{q}_i,t\right) >0
\label{Eq.16}
\end{equation}

With these conditions satisfied the Lagrangian $L^{\prime }$ will have a
minimum in $\stackrel{.}{q_i}=\beta _i\left( q_i,t\right) $ and $\stackrel{..%
}{q}_i=\varphi _i\left( q_i,\stackrel{.}{q}_i,t\right) $ and consequently
the action integral will have a minimum. So, the solutions of the
differential equations will correspond to extremes of the action integral.

{}From the definition of $L^{\prime }$ we have:
\begin{equation}  \label{Eq.17}
L^{\prime }=L\left( q_i,\stackrel{.}{q}_i,\stackrel{..}{q}_i,t\right) -\frac{%
\partial S\left( q_i,\stackrel{.}{q}_i,t\right) }{\partial t}-\frac{\partial
S\left( q_i,\stackrel{.}{q}_i,t\right) }{\partial q_i}\frac{dq_i}{dt}-\frac{%
\partial S\left( q_i,\stackrel{.}{q}_i,t\right) }{\partial \stackrel{.}{q}_i}%
\frac{d\stackrel{.}{q}_i}{dt}
\end{equation}

Using condition (\ref{Eq.15}) we have:
\begin{eqnarray}
&&\left[ L\left( q_i,\stackrel{.}{q}_i,\stackrel{..}{q}_i,t\right) -\frac{%
\partial S\left( q_i,\stackrel{.}{q}_i,t\right) }{\partial t}\right.
\label{Eq.18} \\
&&_{}{} \hskip 2.6cm \left. \left. -\frac{\partial S\left( q_i,\stackrel{.}{q%
}_i,t\right) }{\partial q_i}\stackrel{.}{q}_i-\frac{\partial S\left( q_i,%
\stackrel{.}{q}_i,t\right) }{\partial \stackrel{.}{q}_i}\stackrel{..}{q}%
_i\right] \right| \Sb \stackrel{.}{q}_i=\beta _i  \\ \stackrel{..}{q}%
_i=\varphi _i  \endSb =0  \nonumber \\
&&  \nonumber
\end{eqnarray}
\begin{equation}
\left. \frac{\partial S}{\partial t}\right| \Sb \stackrel{.}{q}_i=\beta _i
\\ \stackrel{..}{q}_i=\varphi _i  \endSb =\left. \left[ L\left( q_i,%
\stackrel{.}{q}_i,\stackrel{..}{q}_i,t\right) -\frac{\partial S\left( q_i,%
\stackrel{.}{q}_i,t\right) }{\partial q_i}\stackrel{.}{q}_i-\frac{\partial
S\left( q_i,\stackrel{.}{q}_i,t\right) }{\partial \stackrel{.}{q}_i}%
\stackrel{..}{q}_i\right] \right| \Sb \stackrel{.}{q}_i=\beta _i  \\
\stackrel{..}{q}_i=\varphi _i  \endSb   \label{Eq.19}
\end{equation}

Since $\stackrel{.}{q}_i=\beta _i$ and $\stackrel{..}{q}_i=\varphi _i$ are
minimum points of $L^{\prime }$ we must have
\begin{equation}  \label{Eq.20}
\left. \frac{\partial L^{\prime }}{\partial \stackrel{..}{q_i}}\right| \Sb
\stackrel{.}{q}_i=\beta _i  \\ \stackrel{..}{q}_i=\varphi _i  \endSb %
=0\Rightarrow \left. \left[ \frac{\partial L}{\partial \stackrel{..}{q}_i}%
-\frac \partial {\partial \stackrel{..}{q}_i}\left( \frac{dS}{dt}\right)
\right] \right| \Sb \stackrel{.}{q}_i=\beta _i  \\ \stackrel{..}{q}%
_i=\varphi _i  \endSb =0,
\end{equation}
\begin{equation}  \label{Eq.21}
\left. \left[ \frac{\partial L}{\partial \stackrel{..}{q}_i}- \frac{\partial
S}{\partial \stackrel{.}{q}_i}\right] \right| \Sb \stackrel{.}{q}_i=\beta _i
\\ \stackrel{..}{q}_i=\varphi _i  \endSb =0,
\end{equation}
or
\begin{equation}  \label{Eq.22}
\left. \frac{\partial S}{\partial \stackrel{.}{q}_i}\right| _{\stackrel{.}{q}%
_i=\beta _i}=\left. \frac{\partial L}{\partial \stackrel{..}{q}_i}\right| \Sb
\stackrel{.}{q}_i=\beta _i  \\ \stackrel{..}{q}_i=\varphi _i  \endSb
\end{equation}

Analogously we must have
\begin{equation}  \label{Eq.23}
\left. \frac{\partial L^{\prime }}{\partial \stackrel{.}{q}_i}\right| \Sb
\stackrel{.}{q}_i=\beta _i  \\ \stackrel{..}{q}_i=\varphi _i  \endSb %
=0\Rightarrow \left. \left[ \frac{\partial L}{\partial \stackrel{.}{q}_i}%
-\frac \partial {\partial \stackrel{.}{q}_i}\left( \frac{dS}{dt}\right)
\right] \right| \Sb \stackrel{.}{q}_i=\beta _i  \\ \stackrel{..}{q}%
_i=\varphi _i  \endSb =0,
\end{equation}
\begin{equation}  \label{Eq.24}
\left. \left[ \frac{\partial L}{\partial \stackrel{.}{q}_i}-\frac \partial
{\partial t}\frac{\partial S}{\partial \stackrel{.}{q}_i}- \frac{\partial S}{%
\partial q_i}-\frac{\partial ^2S}{\partial \stackrel{.}{q}_i\partial q_j}%
\stackrel{.}{q}_j-\frac{\partial ^2S}{\partial \stackrel{.}{q}_i\partial
\stackrel{.}{q}_j}\stackrel{..}{q}_j\right] \right| \Sb \stackrel{.}{q}%
_i=\beta _i  \\ \stackrel{..}{q}_i=\varphi _i  \endSb =0,
\end{equation}
\begin{equation}  \label{Eq.25}
\left. \left[ \frac{\partial L}{\partial \stackrel{.}{q}_i}- \frac{\partial S%
}{\partial q_i}-\frac d{dt}\frac{\partial S}{\partial \stackrel{.}{q}_i}%
\right] \right| \Sb \stackrel{.}{q}_i=\beta _i  \\ \stackrel{..}{q}%
_i=\varphi _i  \endSb =0,
\end{equation}
or
\begin{equation}  \label{Eq.26}
\left. \frac{\partial S}{\partial q_i}\right| _{\stackrel{.}{q}_i=\beta
_i}=\left. \left[ \frac{\partial L}{\partial \stackrel{.}{q}_i}-\frac d{dt}%
\frac{\partial S}{\partial \stackrel{.}{q}_i}\right] \right| \Sb \stackrel{.%
}{q}_i=\beta _i  \\ \stackrel{..}{q}_i=\varphi _i  \endSb
\end{equation}

{}From these results, using the definitions for the conjugated momenta given
by equations (\ref{Eq.3}) and (\ref{Eq.4}) and writing $\stackrel{.}{q}_i=%
\stackrel{\_}{q}_i$, we have from equation (\ref{Eq.19}) that, to obtain an
extreme of the action, we must get a function $S(q_i,\stackrel{.}{q}_i,t)$
such that:
\begin{equation}  \label{Eq.27}
\frac{\partial S}{\partial t}=-H_0
\end{equation}
\begin{equation}  \label{Eq.28}
H_0=p_i\stackrel{\_}{q}_i+\pi _i\stackrel{\stackrel{.}{\_}}{q}_i-L
\end{equation}
\begin{equation}  \label{Eq.29}
p_i=\frac{\partial S}{\partial q_i};\pi _i=\frac{\partial S}{\partial
\stackrel{\_}{q}_i}
\end{equation}

These are the fundamental equations of the equivalent Lagrangian method,
equation (\ref{Eq.27}) being called the Hamilton-Jacobi partial differential
equation, or simply the HJPDE.

\section{Formulation for singular second order Lagrangians}

We consider now the application of the formalism developed in the previous
section to a system with a singular second order Lagrangian. When the
Hessian matrix has a rank $n-R$, $R<n$, the momenta variables will not be
independent variables among themselves. In this case we can choose the order
of accelerations $\stackrel{\stackrel{.}{\_}}{q}_i=\stackrel{..}{q}_i$ in
such a way that the minor of rank $n-R$ in the bottom right corner has
nonvanishing determinant:
\begin{equation}  \label{Eq.30}
\det \left\| \frac{\partial ^2L}{\partial \stackrel{\stackrel{.}{\_}}{q}%
_a\partial \stackrel{\stackrel{.}{\_}}{q}_b}\right\| =\det \left\| \frac{%
\partial \pi _b}{\partial \stackrel{\stackrel{.}{\_}}{q}_a}\right\| \neq
0;a,b=R+1,...,n
\end{equation}

So we can solve the $n-R$ accelerations $\stackrel{\stackrel{.}{\_}}{q}_a$
in terms of the coordinates $\left( q,\stackrel{\_}{q}\right) $, the momenta
$\pi _a$ and the unsolved accelerations $\stackrel{\stackrel{.}{\_}}{q}%
_\alpha $ $(\alpha $$=1,...,R)$ as follow:
\begin{equation}
\stackrel{\stackrel{.}{\_}}{q}_a=f_a\left( q_i,\stackrel{\_}{q}_i,\pi _b,%
\stackrel{\stackrel{.}{\_}}{q}_\alpha \right)  \label{Eq.31}
\end{equation}

Since the momenta $\pi $ are functions of the accelerations $\stackrel{%
\stackrel{.}{\_}}{q}_i$ we can substitute the expressions (\ref{Eq.31}) and
obtain:
\begin{equation}  \label{Eq.32}
\pi _i=g_i\left( q_i,\stackrel{\_}{q}_i,\stackrel{\stackrel{.}{\_}}{q}%
_i\right) =g_i\left( q_i,\stackrel{\_}{q}_i,f_a,\stackrel{\stackrel{.}{\_}}{q%
}_\alpha \right)
\end{equation}
\begin{equation}  \label{Eq.33}
\pi _i=g_i\left( q_i,\stackrel{\_}{q_i},\pi _a,\stackrel{\stackrel{.}{\_}}{q}%
_\alpha \right)
\end{equation}

Since we have $\pi _a\equiv g_a$ the other $n-R$ functions $g_\alpha $ can't
contain the unsolved accelerations $\stackrel{\stackrel{.}{\_}}{q}_\alpha $
or we would be able to solve more of the accelerations $\stackrel{\stackrel{.%
}{\_}}{q}_i$ as functions of the other variables, which contradicts the fact
that the rank of the Hessian matrix is $n-R$. So, we can write for the
momenta $\pi _\alpha $:
\begin{equation}
\pi _\alpha =-H_\alpha ^\pi \left( q_i,\stackrel{\_}{q}_i,p_a,\pi _a\right)
\label{Eq.34}
\end{equation}

We can obtain a similar expression for the momenta $p_\alpha $:
\begin{equation}
p_\alpha =-H_\alpha ^p\left( q_i,\stackrel{\_}{q}_i,p_a,\pi _a\right)
\label{Eq.35}
\end{equation}

Anyway, from a general constraint given by any expression like equation (\ref
{Eq.6}) we can always obtain expressions like equations (\ref{Eq.34}) and (%
\ref{Eq.35}) (see ref.\cite{9}).

The Hamiltonian $H_0$, given by equation (\ref{Eq.28}), becomes
\begin{eqnarray}
H_0 &=&p_a\stackrel{\_}{q}_a+\stackrel{\_}{q}_\alpha \left. p_\alpha \right|
_{p_\beta =-H_\beta ^p}{}_{}{}_{}+\pi _af_a  \label{Eq.36} \\
&&+\stackrel{\stackrel{.}{\_}}{q}_\alpha \left. \pi _\alpha \right| _{\pi
_\beta =-H_\beta ^\pi }{}_{}{}_{}-L\left( q_i,\stackrel{\_}{q_i},\stackrel{%
\stackrel{.}{\_}}{q}_\alpha ,\stackrel{\stackrel{.}{\_}}{q}_a=f_a\right)
\nonumber \\
\nonumber
\end{eqnarray}
where $\alpha ,\beta =1,...,R$; $a=R+1,...,n$. On the other hand we have
\begin{equation}
\frac{\partial H_0}{\partial \stackrel{\stackrel{.}{\_}}{q}_\alpha }=\pi _a%
\frac{\partial f_a}{\partial \stackrel{\stackrel{.}{\_}}{q}_\alpha }+\pi
_\alpha -\frac{\partial L}{\partial \stackrel{\stackrel{.}{\_}}{q}_\alpha }-%
\frac{\partial L}{\partial \stackrel{\stackrel{.}{\_}}{q}_a}\frac{\partial
f_a}{\partial \stackrel{\stackrel{.}{\_}}{q}_\alpha }=0  \label{Eq.37}
\end{equation}
so the Hamiltonian $H_0$ does not depend explicitly upon the accelerations $%
\stackrel{\stackrel{.}{\_}}{q}_\alpha $.

{}From this point we will adopt the following notation: the coordinates $t$
and $q_\alpha $ will be called $t_0$ and $t_\alpha $, respectively, and $%
\stackrel{\_}{q}_\alpha $ will be called $\stackrel{\_}{t}_\alpha $. The
momenta $p_\alpha $ and $\pi _\alpha $ will be called $P_\alpha ^p$ and $%
P_\alpha ^\pi $, respectively, and the momentum $P_0$ will be defined as:
\begin{equation}
P_0=\frac{\partial S}{\partial t}  \label{Eq.38}
\end{equation}

Then, to obtain an extreme of the action integral, we must find a function $%
S\left( q_i,\stackrel{.}{q_i},t\right) $ that satisfies the following set of
HJPDE:
\begin{equation}
H_0^{\prime }=P_0+H_0\left( t_0,t_\alpha ,\stackrel{\_}{t}_\alpha ;q_a,%
\stackrel{\_}{q}_a;p_a=\frac{\partial S}{\partial q_a};\pi _a=\frac{\partial
S}{\partial \stackrel{\_}{q}_a}\right) =0  \label{Eq.39}
\end{equation}
\begin{equation}
H_\alpha ^{\prime p}=P_\alpha ^p+H_\alpha ^p\left( t_0,t_\alpha ,\stackrel{\_%
}{t}_\alpha ;q_a,\stackrel{\_}{q}_a;p_a=\frac{\partial S}{\partial q_a};\pi
_a=\frac{\partial S}{\partial \stackrel{\_}{q}_a}\right) =0  \label{Eq.40}
\end{equation}
\begin{equation}
H_\alpha ^{\prime \pi }=P_\alpha ^\pi +H_\alpha ^\pi \left( t_0,t_\alpha ,%
\stackrel{\_}{t}_\alpha ;q_a,\stackrel{\_}{q}_a;p_a=\frac{\partial S}{%
\partial q_a};\pi _a=\frac{\partial S}{\partial \stackrel{\_}{q}_a}\right) =0
\label{Eq.41}
\end{equation}

{}From the definition above and equation (\ref{Eq.36}) we have
\begin{equation}  \label{Eq.42}
\frac{\partial H_0^{\prime }}{\partial \pi _b}=-\frac{\partial L }{\partial
\stackrel{\stackrel{.}{\_}}{q}_a}\frac{\partial f_a}{\partial \pi _b}-%
\stackrel{\stackrel{.}{\_}}{q}_\alpha \frac{\partial H_\alpha ^\pi }{%
\partial \pi _b}-\stackrel{\_}{q}_\alpha \frac{\partial H_\alpha ^p}{%
\partial \pi _b}+\pi _a\frac{\partial f_a}{\partial \pi _b}+\stackrel{%
\stackrel{.}{\_}}{q}_b
\end{equation}
\begin{equation}  \label{Eq.43}
\frac{\partial H_0^{\prime }}{\partial \pi _b}=\stackrel{\stackrel{.}{\_}}{q}%
_b-\stackrel{\stackrel{.}{\_}}{q}_\alpha \frac{\partial H_\alpha ^\pi }{%
\partial \pi _b}-\stackrel{\_}{q}_\alpha \frac{\partial H_\alpha ^p}{%
\partial \pi _b}
\end{equation}
and
\begin{equation}  \label{Eq.44}
\frac{\partial H_0^{\prime }}{\partial p_b}=-\stackrel{\stackrel{.}{\_}}{q}%
_\alpha \frac{\partial H_\alpha ^\pi }{\partial p_b}-\stackrel{\_}{q}_\alpha
\frac{\partial H_\alpha ^p}{\partial p_b}+\stackrel{\_}{q}_b=\stackrel{\_}{q}%
_b-\stackrel{\stackrel{.}{\_}}{q}_\alpha \frac{\partial H_\alpha ^\pi }{%
\partial p_b}-\stackrel{\_}{q}_\alpha \frac{\partial H_\alpha ^p}{\partial
p_b}
\end{equation}
where $\alpha =1,...,R$; $a,b=R+1,...,n$.

Remembering that $\stackrel{.}{q}_i=\stackrel{\_}{q}_i$ and multiplying by $%
dt=dt_0$ we have from equations (\ref{Eq.43}) and (\ref{Eq.44})
\begin{equation}  \label{Eq.45}
d\stackrel{\_}{q}_b=\frac{\partial H_0^{\prime }}{\partial \pi _b}dt_0+\frac{%
\partial H_\alpha ^{\prime p}}{\partial \pi _b}dt_\alpha + \frac{\partial
H_\alpha ^{\prime \pi }}{\partial \pi _b}d\stackrel{\_}{t}_\alpha
\end{equation}
\begin{equation}  \label{Eq.46}
dq_b=\frac{\partial H_0^{\prime }}{\partial p_b}dt_0+\frac{\partial H_\alpha
^{\prime p}}{\partial p_b}dt_\alpha +\frac{\partial H_\alpha ^{\prime \pi }}{%
\partial p_b}d\stackrel{\_}{t}_\alpha
\end{equation}
and also
\begin{equation}  \label{Eq.47}
dq_\alpha =\frac{\partial H_0^{\prime }}{\partial P_\alpha ^p}dt_0+\frac{%
\partial H_\beta ^{\prime p}}{\partial P_\alpha ^p}dt_\beta + \frac{\partial
H_\beta ^{\prime \pi }}{\partial P_\alpha ^p}d\stackrel{\_}{t}_\beta =\frac{%
\partial H_\beta ^{\prime p}}{\partial P_\alpha ^p}dt_\beta =\delta _{\alpha
\beta }dt_\beta =dt_\alpha
\end{equation}
\begin{equation}  \label{Eq.48}
d\stackrel{\_}{q}_\alpha =\frac{\partial H_0^{\prime }}{\partial P_\alpha
^\pi }dt_0+\frac{\partial H_\beta ^{\prime p}}{\partial P_\alpha ^\pi }%
dt_\beta +\frac{\partial H_\beta ^{\prime \pi }}{\partial P_\alpha ^\pi }d%
\stackrel{\_}{t}_\beta =\frac{\partial H_\beta ^{\prime p}}{\partial
P_\alpha ^\pi }dt_\beta =\delta _{\alpha \beta }d\stackrel{\_}{t}_\beta =d%
\stackrel{\_}{t}_\alpha
\end{equation}
\begin{equation}  \label{Eq.49}
dq_0=dt=\frac{\partial H_0^{\prime }}{\partial P_0}dt_0+\frac{\partial
H_\beta ^{\prime p}}{\partial P_0}dt_\beta +\frac{\partial H_\beta ^{\prime
\pi }}{\partial P_0}d\stackrel{\_}{t}_\beta =\frac{\partial H_0^{\prime }}{%
\partial P_0}dt_0=dt_0
\end{equation}
for $\beta =1,...,R$; so we can write the equations (\ref{Eq.45}) and (\ref
{Eq.46}) as:
\begin{equation}  \label{Eq.50}
d\stackrel{\_}{q}_i=\frac{\partial H_0^{\prime }}{\partial \pi _i}dt_0+\frac{%
\partial H_\alpha ^{\prime p}}{\partial \pi _i}dt_\alpha + \frac{\partial
H_\alpha ^{\prime \pi }}{\partial \pi _i}d\stackrel{\_}{t}_\alpha ;i=1,...,n
\end{equation}
\begin{equation}  \label{Eq.51}
dq_i=\frac{\partial H_0^{\prime }}{\partial p_i}dt_0+\frac{\partial H_\alpha
^{\prime p}}{\partial p_i}dt_\alpha +\frac{\partial H_\alpha ^{\prime \pi }}{%
\partial p_i}d\stackrel{\_}{t}_\alpha ;i=1,...,n
\end{equation}

If we consider that we have a solution $S\left( q_i,\stackrel{.}{q_i}%
,t\right) $ of the set of HJPDE given by equations (\ref{Eq.39}), (\ref
{Eq.40}) and (\ref{Eq.41}) then, differentiating those equations with
respect to $\stackrel{\_}{q}_i$, we obtain
\begin{equation}
\frac{\partial H_0^{\prime }}{\partial \stackrel{\_}{q}_i}+\frac{\partial
H_0^{\prime }}{\partial P_0}\frac{\partial ^2S}{\partial t\partial \stackrel{%
\_}{q}_i}+\frac{\partial H_0^{\prime }}{\partial p_a}\frac{\partial ^2S}{%
\partial q_a\partial \stackrel{\_}{q}_i}+\frac{\partial H_0^{\prime }}{%
\partial \pi _a}\frac{\partial ^2S}{\partial \stackrel{\_}{q}_a\partial
\stackrel{\_}{q}_i}=0  \label{Eq.52}
\end{equation}
\begin{equation}
\frac{\partial H_\alpha ^{\prime p}}{\partial \stackrel{\_}{q}_i}+\frac{%
\partial H_\alpha ^{\prime p}}{\partial P_\beta ^p}\frac{\partial ^2S}{%
\partial t_\beta \partial \stackrel{\_}{q}_i}+\frac{\partial H_\alpha
^{\prime p}}{\partial p_a}\frac{\partial ^2S}{\partial q_a\partial \stackrel{%
\_}{q}_i}+\frac{\partial H_\alpha ^{\prime p}}{\partial \pi _a}\frac{%
\partial ^2S}{\partial \stackrel{\_}{q}_a\partial \stackrel{\_}{q}_i}=0
\label{Eq.53}
\end{equation}
\begin{equation}
\frac{\partial H_\alpha ^{\prime \pi }}{\partial \stackrel{\_}{q}_i}+\frac{%
\partial H_\alpha ^{\prime \pi }}{\partial P_\beta ^\pi }\frac{\partial ^2S}{%
\partial \stackrel{\_}{t}_\beta \partial \stackrel{\_}{q}_i}+\frac{\partial
H_\alpha ^{\prime \pi }}{\partial p_a}\frac{\partial ^2S}{\partial
q_a\partial \stackrel{\_}{q}_i}+\frac{\partial H_\alpha ^{\prime \pi }}{%
\partial \pi _a}\frac{\partial ^2S}{\partial \stackrel{\_}{q}_a\partial
\stackrel{\_}{q}_i}=0  \label{Eq.54}
\end{equation}
whereas for $q_i$ we get
\begin{equation}
\frac{\partial H_0^{\prime }}{\partial q_i}+\frac{\partial H_0^{\prime }}{%
\partial P_0}\frac{\partial ^2S}{\partial t\partial q_i}+\frac{\partial
H_0^{\prime }}{\partial p_a}\frac{\partial ^2S}{\partial q_a\partial q_i}+%
\frac{\partial H_0^{\prime }}{\partial \pi _a}\frac{\partial ^2S}{\partial
\stackrel{\_}{q}_a\partial q_i}=0  \label{Eq.55}
\end{equation}
\begin{equation}
\frac{\partial H_\alpha ^{\prime p}}{\partial q_i}+\frac{\partial H_\alpha
^{\prime p}}{\partial P_\beta ^p}\frac{\partial ^2S}{\partial t_\beta
\partial q_i}+\frac{\partial H_\alpha ^{\prime p}}{\partial p_a}\frac{%
\partial ^2S}{\partial q_a\partial q_i}+\frac{\partial H_\alpha ^{\prime p}}{%
\partial \pi _a}\frac{\partial ^2S}{\partial \stackrel{\_}{q}_a\partial q_i}%
=0  \label{Eq.56}
\end{equation}
\begin{equation}
\frac{\partial H_\alpha ^{\prime \pi }}{\partial q_i}+\frac{\partial
H_\alpha ^{\prime \pi }}{\partial P_\beta ^\pi }\frac{\partial ^2S}{\partial
\stackrel{\_}{t}_\beta \partial q_i}+\frac{\partial H_\alpha ^{\prime \pi }}{%
\partial p_a}\frac{\partial ^2S}{\partial q_a\partial q_i}+\frac{\partial
H_\alpha ^{\prime \pi }}{\partial \pi _a}\frac{\partial ^2S}{\partial
\stackrel{\_}{q}_a\partial q_i}=0  \label{Eq.57}
\end{equation}
and for $t_0$ we have:
\begin{equation}
\frac{\partial H_0^{\prime }}{\partial t_0}+\frac{\partial H_0^{\prime }}{%
\partial P_0}\frac{\partial ^2S}{\partial t\partial t_0}+\frac{\partial
H_0^{\prime }}{\partial p_a}\frac{\partial ^2S}{\partial q_a\partial t_0}+%
\frac{\partial H_0^{\prime }}{\partial \pi _a}\frac{\partial ^2S}{\partial
\stackrel{\_}{q}_a\partial t_0}=0  \label{Eq.58}
\end{equation}
\begin{equation}
\frac{\partial H_\alpha ^{\prime p}}{\partial t_0}+\frac{\partial H_\alpha
^{\prime p}}{\partial P_\beta ^p}\frac{\partial ^2S}{\partial t_\beta
\partial t_0}+\frac{\partial H_\alpha ^{\prime p}}{\partial p_a}\frac{%
\partial ^2S}{\partial q_a\partial t_0}+\frac{\partial H_\alpha ^{\prime p}}{%
\partial \pi _a}\frac{\partial ^2S}{\partial \stackrel{\_}{q}_a\partial t_0}%
=0  \label{Eq.59}
\end{equation}
\begin{equation}
\frac{\partial H_\alpha ^{\prime \pi }}{\partial t_0}+\frac{\partial
H_\alpha ^{\prime \pi }}{\partial P_\beta ^\pi }\frac{\partial ^2S}{\partial
\stackrel{\_}{t}_\beta \partial t_0}+\frac{\partial H_\alpha ^{\prime \pi }}{%
\partial p_a}\frac{\partial ^2S}{\partial q_a\partial t_0}+\frac{\partial
H_\alpha ^{\prime \pi }}{\partial \pi _a}\frac{\partial ^2S}{\partial
\stackrel{\_}{q}_a\partial t_0}=0  \label{Eq.60}
\end{equation}

Making $Z=S\left( q_i,\stackrel{.}{q_i},t\right) $and using the momenta
definitions together with equations (\ref{Eq.50}) and (\ref{Eq.51}) we have
\begin{equation}
dZ=\frac{\partial S}{\partial t}dt_0+\frac{\partial S}{\partial t_\alpha }%
dt_\alpha +\frac{\partial S}{\partial \stackrel{\_}{t}_\alpha }d\stackrel{\_%
}{t}_\alpha +\frac{\partial S}{\partial q_a}dq_a+\frac{\partial S}{\partial
\stackrel{\_}{q}_a}d\stackrel{\_}{q}_a,  \label{Eq.61}
\end{equation}

\begin{eqnarray}
dZ &=&-H_0dt_0-H_\alpha ^pdt_\alpha -H_\alpha ^\pi d\stackrel{\_}{t}_\alpha
\label{Eq.62} \\
&&+p_a\left( \frac{\partial H_0^{\prime }}{\partial p_a}dt_0+\frac{\partial
H_\alpha ^{\prime p}}{\partial p_a}dt_\alpha +\frac{\partial H_\alpha
^{\prime \pi }}{\partial p_a}d\stackrel{\_}{t}_\alpha \right)  \nonumber \\
&&+\pi _a\left( \frac{\partial H_0^{\prime }}{\partial \pi _a}dt_0+\frac{%
\partial H_\alpha ^{\prime p}}{\partial \pi _a}dt_\alpha +\frac{\partial
H_\alpha ^{\prime \pi }}{\partial \pi _a}d\stackrel{\_}{t}_\alpha \right) ,
\nonumber \\
&&  \nonumber
\end{eqnarray}
\begin{eqnarray}
dZ &=&\left( -H_0+p_a\frac{\partial H_0^{\prime }}{\partial p_a}+\pi _a\frac{%
\partial H_0^{\prime }}{\partial \pi _a}\right) dt_0  \label{Eq.63} \\
&&+\left( -H_\alpha ^p+p_a\frac{\partial H_\alpha ^{\prime p}}{\partial p_a}%
+\pi _a\frac{\partial H_\alpha ^{\prime p}}{\partial \pi _a}\right) dt_\alpha
\nonumber \\
&&+\left( -H_\alpha ^\pi +p_a\frac{\partial H_\alpha ^{\prime \pi }}{%
\partial p_a}+\pi _a+\frac{\partial H_\alpha ^{\prime \pi }}{\partial \pi _a}%
\right) d\stackrel{\_}{t}_\alpha  \nonumber \\
&&  \nonumber
\end{eqnarray}
and, from momenta definitions:
\begin{equation}
dP_0=\frac{\partial ^2S}{\partial ^2t}dt_0+\frac{\partial ^2S}{\partial
t\partial t_\alpha }dt_\alpha +\frac{\partial ^2S}{\partial t\partial q_a}%
dq_a+\frac{\partial ^2S}{\partial t\partial \stackrel{\_}{t}_\alpha }d%
\stackrel{\_}{t}_\alpha +\frac{\partial ^2S}{\partial t\partial \stackrel{\_%
}{q}_a}d\stackrel{\_}{q}_a  \label{Eq.64}
\end{equation}
\begin{equation}
dp_i=\frac{\partial ^2S}{\partial q_i\partial t}dt_0+\frac{\partial ^2S}{%
\partial q_i\partial t_\alpha }dt_\alpha +\frac{\partial ^2S}{\partial
q_i\partial q_a}dq_a+\frac{\partial ^2S}{\partial q_i\partial \stackrel{\_}{t%
}_\alpha }d\stackrel{\_}{t}_\alpha +\frac{\partial ^2S}{\partial q_i\partial
\stackrel{\_}{q}_a}d\stackrel{\_}{q}_a  \label{Eq.65}
\end{equation}
\begin{eqnarray}
d\pi _i &=&\frac{\partial ^2S}{\partial \stackrel{\_}{q}_i\partial t}dt_0+%
\frac{\partial ^2S}{\partial \stackrel{\_}{q}_i\partial t_\alpha }dt_\alpha +%
\frac{\partial ^2S}{\partial \stackrel{\_}{q}_i\partial q_a}dq_a
\label{Eq.66} \\
&&+\frac{\partial ^2S}{\partial \stackrel{\_}{q}_i\partial \stackrel{\_}{t}%
_\alpha }d\stackrel{\_}{t}_\alpha +\frac{\partial ^2S}{\partial \stackrel{\_%
}{q}_i\partial \stackrel{\_}{q}_a}d\stackrel{\_}{q}_a  \nonumber
\end{eqnarray}

Now, multiplying equations (\ref{Eq.58}) by $dt_0$, contracting equations (%
\ref{Eq.59}) and (\ref{Eq.60}) with $dt_\alpha $ and $d\stackrel{\_}{t}%
_\alpha $ (respectively) and adding them all to equation (\ref{Eq.64}) we
get:

\begin{eqnarray}
dP_0&+&\frac{\partial H_0^{\prime }}{\partial t_0}dt_0+\frac{\partial
H_\alpha ^{\prime p}}{\partial t_0}dt_\alpha +\frac{\partial H_\alpha
^{\prime \pi }}{\partial t_0}d\stackrel{\_}{t }_\alpha=  \label{Eq.67} \\
&&=\frac{\partial ^2S}{\partial t_0\partial q_a}\left( dq_a-\frac{\partial
H_0^{\prime }}{\partial p_a}dt_0-\frac{\partial H_\alpha ^{\prime p}}{%
\partial p_a}dt_\alpha -\frac{\partial H_\alpha ^{\prime \pi }}{\partial p_a}%
d\stackrel{\_}{t }_\alpha \right)  \nonumber \\
&& + \frac{\partial ^2S}{\partial t_0\partial \stackrel{\_}{q}_a}\left( d%
\stackrel{\_}{q}_a-\frac{\partial H_0^{\prime }}{\partial \pi _a}dt_0-\frac{%
\partial H_\alpha ^{\prime p}}{\partial \pi _a}dt_\alpha -\frac{\partial
H_\alpha ^{\prime \pi }}{\partial \pi _a}d\stackrel{\_}{t }_\alpha \right)
\nonumber \\
\nonumber
\end{eqnarray}

In a similar way, using the same steps with equations (\ref{Eq.55}-\ref
{Eq.57}) and (\ref{Eq.65}) we obtain:

\begin{eqnarray}
dp_i&+&\frac{\partial H_0^{\prime }}{\partial q_i}dt_0+\frac{\partial
H_\alpha ^{\prime p}}{\partial q_i}dt_\alpha +\frac{\partial H_\alpha
^{\prime \pi }}{\partial q_i}d\stackrel{\_}{t }_\alpha =  \label{Eq.68} \\
&&=\frac{\partial ^2S}{\partial q_i\partial q_a}\left( dq_a-\frac{\partial
H_0^{\prime }}{\partial p_a}dt_0-\frac{\partial H_\alpha ^{\prime p}}{%
\partial p_a}dt_\alpha -\frac{\partial H_\alpha ^{\prime \pi }}{\partial p_a}%
d\stackrel{\_}{t }_\alpha \right)  \nonumber \\
&&+ \frac{\partial ^2S}{\partial q_i\partial \stackrel{\_}{q}_a}\left( d%
\stackrel{\_}{q}_a-\frac{\partial H_0^{\prime }}{\partial \pi _a}dt_0-\frac{%
\partial H_\alpha ^{\prime p}}{\partial \pi _a}dt_\alpha -\frac{\partial
H_\alpha ^{\prime \pi }}{\partial \pi _a}d\stackrel{\_}{t }_\alpha \right)
\nonumber \\
\nonumber
\end{eqnarray}

And, finally, using equations (\ref{Eq.52}-\ref{Eq.54}) and (\ref{Eq.66}) we
have:

\begin{eqnarray}
d\pi _i&+&\frac{\partial H_0^{\prime }}{\partial \stackrel{\_}{q}_i}dt_0+%
\frac{\partial H_\alpha ^{\prime p}}{\partial \stackrel{\_}{q}_i}dt_\alpha +%
\frac{\partial H_\alpha ^{\prime \pi }}{\partial \stackrel{\_}{q}_i}d%
\stackrel{\_}{t }_\alpha=  \label{Eq.69} \\
&&=\frac{\partial ^2S}{\partial \stackrel{\_}{q}_i\partial q_a}\left( dq_a-%
\frac{\partial H_0^{\prime }}{\partial p_a}dt_0-\frac{\partial H_\alpha
^{\prime p}}{\partial p_a}dt_\alpha -\frac{\partial H_\alpha ^{\prime \pi }}{%
\partial p_a}d\stackrel{\_}{t }_\alpha \right)  \nonumber \\
&&+\frac{\partial ^2S}{\partial \stackrel{\_}{q}_i\partial \stackrel{\_}{q}_a%
}\left( d\stackrel{\_}{q}_a-\frac{\partial H_0^{\prime }}{\partial \pi _a}%
dt_0-\frac{\partial H_\alpha ^{\prime p}}{\partial \pi _a}dt_\alpha -\frac{%
\partial H_\alpha ^{\prime \pi }}{\partial \pi _a}d\stackrel{\_}{t }_\alpha
\right)  \nonumber \\
\nonumber
\end{eqnarray}

If the total differential equations given by (\ref{Eq.45}) and (\ref{Eq.46})
are valid the equations (\ref{Eq.67}), (\ref{Eq.68}) and (\ref{Eq.69})
become:
\begin{equation}  \label{Eq.70}
dP_0=\frac{\partial H_0^{\prime }}{\partial t_0}dt_0-\frac{\partial H_\alpha
^{\prime p}}{\partial t_0}dt_\alpha -\frac{\partial H_\alpha ^{\prime \pi }}{%
\partial t_0}d\stackrel{\_}{t}_\alpha
\end{equation}
\begin{equation}  \label{Eq.71}
dp_i=-\frac{\partial H_0^{\prime }}{\partial q_i}dt_0-\frac{\partial
H_\alpha ^{\prime p}}{\partial q_i}dt_\alpha -\frac{\partial H_\alpha
^{\prime \pi }}{\partial q_i}d\stackrel{\_}{t}_\alpha
\end{equation}
\begin{equation}  \label{Eq.72}
d\pi _i-\frac{\partial H_0^{\prime }}{\partial \stackrel{\_}{q}_i}dt_0-\frac{%
\partial H_\alpha ^{\prime p}}{\partial \stackrel{\_}{q}_i}dt_\alpha -\frac{%
\partial H_\alpha ^{\prime \pi }}{\partial \stackrel{\_}{q}_i}d\stackrel{\_}{%
t}_\alpha
\end{equation}

These equations together with equations (\ref{Eq.49}), (\ref{Eq.50}), (\ref
{Eq.51}) and (\ref{Eq.63}) are the total differential equations for the
characteristics curves and, if they form a completely integrable set, their
simultaneous solutions determine $S\left( q_i,\stackrel{\_}{q}_i,t_0\right) $
uniquely by the initial conditions.

\section{Example: Podolsky generalized electrodyna\-mics}

In this section we will consider a continuous system with Lagrangian density
dependent on the dynamical field variables and its derivatives upon second
order: ${\cal L=L}\left( \psi ,\partial \psi ,\partial ^2\psi \right) $. We
adopt the metric $\eta _{\mu \nu }=diag(+1,-1,-1,-1)$ with Greek indices
running from 0 to 3 while Latin indices run from 1 to 3. As stated
previously, the generalization of the forma\-lism presented in Sections 4
and 5 is straightforward, being necessary only to consider that the
Euler-Lagrange equations of motion are now given by
\begin{equation}
\frac{\partial {\cal L}}{\partial \psi ^a}-\partial _\mu \left( \frac{%
\partial {\cal L}}{\partial \left( \partial _\mu \psi ^a\right) }\right)
+\partial _\mu \partial _\nu \left( \frac{\partial {\cal L}}{\partial \left(
\partial _\mu \partial _\nu \psi ^a\right) }\right) =0  \label{Eq.73}
\end{equation}
and that the momenta, conjugated respectively to $\stackrel{.}{\psi }^a$ and
$\stackrel{..}{\psi }^a$, are:
\begin{equation}
p_a=\frac{\partial {\cal L}}{\partial \stackrel{.}{\psi }^a}-2\partial
_k\left( \frac{\partial {\cal L}}{\partial \left( \partial _k\stackrel{.}{%
\psi }^a\right) }\right) -\partial _0\left( \frac{\partial {\cal L}}{%
\partial \stackrel{..}{\psi }^a}\right)  \label{Eq.74}
\end{equation}
\begin{equation}
\pi _a=\frac{\partial {\cal L}}{\partial \stackrel{..}{\psi }^a}
\label{Eq.75}
\end{equation}

The Hessian matrix is now:
\begin{equation}  \label{Eq.76}
H_{ab}=\frac{\partial ^2{\cal L}}{\partial \stackrel{..}{\psi }^a\partial
\stackrel{..}{\psi }^b}
\end{equation}

With these modifications we consider the case of Podolsky electrodynamics
which is based on the following Lagrangian
\begin{equation}  \label{Eq.77}
{\cal L}=-\frac 14F_{\mu \nu }F^{\mu \nu }+a^2\partial _\lambda F^{\alpha
\lambda }\partial _\rho F_\alpha \,^\rho
\end{equation}
were $F_{\mu \nu }=\partial _\mu A_\nu -\partial _\nu A_\mu $.

An analysis of the Hamiltonian formalism for this theory was carried out in
ref.\cite{10} and we compare some of the results presented there with the
formalism developed here. The Euler-Lagrange equations are
\begin{equation}
\left( 1+2a^2\Box \right) \partial _\rho F_\alpha \,^\rho =0  \label{Eq.78}
\end{equation}
with our dynamical variables chosen as $A^\mu $ and $\stackrel{\_}{A}^\mu =%
\stackrel{.}{A}^\mu $. The conjugated momenta given by definitions (\ref
{Eq.74}) and (\ref{Eq.75}) are:
\begin{equation}
p_\mu =-F_{0\mu }-2a^2\left( \partial _k\partial _\lambda F^{0\lambda
}\delta ^k\,_\mu -\partial _0\partial _\lambda F_\mu \,^\lambda \right)
\label{Eq.79}
\end{equation}
\begin{equation}
\pi _\mu =2a^2\left( \partial _\lambda F^{0\lambda }\delta ^0\,_\mu
-\partial _\lambda F_\mu \,^\lambda \right)  \label{Eq.80}
\end{equation}

The primary constraints are:
\begin{equation}  \label{Eq.81}
\Phi _1=\pi _0\approx 0
\end{equation}
\begin{equation}  \label{Eq.82}
\Phi _2=p_0-\partial ^k\pi _k\approx 0
\end{equation}

Using the definition of $\pi $ we can write the accelerations $\stackrel{%
\stackrel{.}{\_}}{A}^i$ as:
\begin{equation}  \label{Eq.83}
\stackrel{\stackrel{.}{\_}}{A}^i=\frac 1{2a^2}\pi ^i+\partial
_kF^{ik}+\partial ^i\stackrel{\_}{A}_0
\end{equation}

The canonical Hamiltonian is given by:
\begin{equation}  \label{Eq.84}
H_c=\int d^3x\left[ p_\mu \stackrel{\_}{A}^\mu +\pi _\mu \stackrel{\stackrel{%
.}{\_}}{A}^\mu -{\cal L}\right]
\end{equation}

Using equation (\ref{Eq.83}) we get:

\begin{eqnarray}
&&H_c=\int d^3x\left[ \stackrel{\_}{A}^0\partial^i\pi_i+p_i\stackrel{\_}{A}%
^i+\frac 1{4a^2}\pi _i\pi ^i+\pi _i\partial _kF^{ik}+\pi _i\partial ^i%
\stackrel{\_}{A}_0 +\frac 14F_{\mu \nu }F^{\mu \nu } \right.  \label{Eq.85}
\\
&& \left. +\frac 12\left( \stackrel{\_}{A}_i-\partial _iA_0\right) \left(
\stackrel{\_}{A}^i-\partial ^iA_0\right) -a^2\left( \partial _k\stackrel{\_}{%
A}^k-\partial _k\partial ^kA_0\right) \left( \partial _i\stackrel{\_}{A}%
^i-\partial _i\partial ^iA_0\right) \right]  \nonumber \\
\nonumber
\end{eqnarray}

According to Dirac's formalism the total Hamiltonian is:
\begin{equation}  \label{Eq.86}
H_T=H_C+\int d^3x\left( C_1\left( x\right) \Phi _1+C_2\left( x\right) \Phi
_2\right)
\end{equation}

The consistency conditions result in:
\begin{equation}  \label{Eq.87}
\stackrel{.}{\Phi }_1=\left\{ \Phi _1,H_T\right\} \approx 0
\end{equation}
\begin{equation}  \label{Eq.88}
\stackrel{.}{\Phi }_2=\left\{ \Phi _2,H_T\right\} =\partial ^kp_k\approx 0
\end{equation}

So we have a secondary constraint given by
\begin{equation}  \label{Eq.89}
\Phi _3=\partial ^kp_k\approx 0
\end{equation}
and the consistency condition results in $\stackrel{.}{\Phi }_3=\left\{ \Phi
_3,H_T\right\} \approx 0$. All constraints are first class so the extended
Hamiltonian is:
\begin{equation}  \label{Eq.90}
H_E=H_C+\int d^3x\left( C_1\left( x\right) \Phi _1+C_2\left( x\right) \Phi
_2+C_3\left( x\right) \Phi _3\right)
\end{equation}

The equations of motion for the dynamical variables, given by $\stackrel{.}{A%
}^\alpha =\left\{ A^\alpha ,H_E\right\} $, are:
\begin{equation}
\stackrel{.}{A}^0=\stackrel{\_}{A}^0+C_2;\;\stackrel{.}{A}^i=\stackrel{\_}{A}%
^i-\partial ^iC_3  \label{Eq.91}
\end{equation}

This simply means that $\stackrel{\_}{A}^\alpha $ is defined as $\stackrel{.%
}{A}^\alpha $ plus additive arbitrary functions. Besides, $\stackrel{%
\stackrel{.}{\_}}{A}^\alpha =\left\{ \stackrel{\_}{A}^\alpha ,H_E\right\} $
gives
\begin{equation}
\stackrel{\stackrel{.}{\_}}{A}^0=C_1;_{}{}_{}{}_{}\stackrel{\stackrel{.}{\_}%
}{A}^i=\frac 1{2a^2}\pi ^i+\partial _kF^{ik}+\partial ^i\stackrel{\_}{A}_0,
\label{Eq.91b}
\end{equation}
which mean that both $\stackrel{\_}{A}^0$ and $A^0$ are arbitrary while we
obtained again equation (\ref{Eq.83}).

For the momenta variables $\stackrel{.}{\pi }_i$$=\left\{ \pi _i,H_E\right\}
$ and $\stackrel{.}{p}_\alpha $$=\left\{ p_\alpha ,H_E\right\} $ give:
\begin{equation}
\stackrel{.}{\pi }_i=-F_{0i}-2a^2\partial _i\partial _kF_0\,^k-p_i
\label{Eq.92}
\end{equation}
\begin{equation}
\stackrel{.}{p}_0=-\partial _iF^{0i}-2a^2\partial ^i\partial _i\partial
_kF_0\,^k  \label{Eq.93}
\end{equation}
\begin{equation}
\stackrel{.}{p}_i=-\partial _i\partial ^k\pi _k+\partial _k\partial ^k\pi
_i-\partial _kF^k\,_i  \label{Eq.94}
\end{equation}

Equation (\ref{Eq.92}) is the definition of $p_i$ given by equation (\ref
{Eq.79}) and together with (\ref{Eq.93}) it gives constraint $\Phi _3$.

Now, using Hamilton-Jacobi formalism we have:
\begin{equation}  \label{Eq.95}
H_0^{\prime }=H_C+P_0;_{}{}_{}P_0=\frac{\partial S}{\partial t}
\end{equation}
\begin{equation}  \label{Eq.96}
H_1^{\prime }=\pi _0;_{}{}_{}H_2^{\prime }=p_0-\partial ^k\pi _k
\end{equation}

The total differential equation for $A^i$ is
\begin{equation}  \label{Eq.96b}
dA^i=\frac{\partial H_0^{\prime }}{\partial p_i}dt+\frac{\partial
H_1^{\prime }}{\partial p_i}d\stackrel{\_}{A}_0+\frac{\partial H_2^{\prime }%
}{\partial p_i}dA_0
\end{equation}
\begin{equation}  \label{Eq.97}
dA^i=\frac{\partial H_0^{\prime }}{\partial p_i}dt=\frac{\partial H_C}{%
\partial p_i}dt\Rightarrow dA^i=\stackrel{\_}{A}^idt
\end{equation}
which is completely equivalent to equation (\ref{Eq.91}) since $C_3$ is
arbitrary. For $\stackrel{\_}{A}^i$ we have:
\begin{equation}  \label{Eq.98}
d\stackrel{\_}{A}^i=\frac{\partial H_0^{\prime }}{\partial \pi _i}dt+\frac{%
\partial H_1^{\prime }}{\partial \pi _i}d\stackrel{\_}{A}_0+ \frac{\partial
H_2^{\prime }}{\partial \pi _i}dA_0=\frac{\partial H_0^{\prime }}{\partial
\pi _i}dt=\frac{\partial H_C}{\partial \pi _i}dt
\end{equation}
\begin{equation}  \label{Eq.99}
d\stackrel{\_}{A}^i=\left( \frac 1{2a^2}\pi ^i+\partial _kF^{ik}+\partial ^i%
\stackrel{\_}{A}_0\right) dt
\end{equation}

Again we have a result in agreement with Dirac's method result given in (\ref
{Eq.91b}).

For the momenta $p_i$ and $p_0$ we have
\begin{equation}  \label{100}
dp^i=-\frac{\partial H_0^{\prime }}{\partial A_i}dt-\frac{\partial
H_1^{\prime }}{\partial A_i}d\stackrel{\_}{A}_0-\frac{\partial H_2^{\prime }%
}{\partial A_i}dA_0=-\frac{\partial H_0^{\prime }}{\partial A_i}dt=-\frac{%
\partial H_C}{\partial A_i}dt
\end{equation}
\begin{equation}  \label{Eq.101}
dp^i=-\int d^3x\left[ \pi _j\partial _k\left( \frac{\partial F^{jk}}{%
\partial A_i}\right) -\frac 12F^{jn}\frac{\partial F_{jn}}{\partial A_i}%
\right] dt
\end{equation}
\begin{equation}  \label{Eq.102}
dp^i=\left[ -\partial ^i\partial ^k\pi _k+\partial _k\partial ^k\pi
^i-\partial _kF^{ki}\right] dt
\end{equation}
and
\begin{equation}  \label{Eq.103}
dp^o=-\frac{\partial H_0^{\prime }}{\partial A_0}dt-\frac{\partial
H_1^{\prime }}{\partial A_0}d\stackrel{\_}{A}_0-\frac{\partial H_2^{\prime }%
}{\partial A_0}dA_0=-\frac{\partial H_C}{\partial A_0}dt
\end{equation}

\begin{eqnarray}
dp^0&=&-\int d^3x\left[ \left( \stackrel{\_}{A}^i-\partial ^iA_0\right)
\frac{\partial \left( \stackrel{\_}{A}_i-\partial _iA_0\right) }{\partial A_0%
} \right.  \label{Eq.104} \\
&& \left. -2a^2\left( \partial _i\stackrel{\_}{A}^i-\partial _i\partial
^iA_0\right) \frac{\partial \left( \partial _k\stackrel{\_}{A}^k-\partial
_k\partial ^kA_0\right) }{\partial A_0}\right] dt  \nonumber \\
\nonumber
\end{eqnarray}
\begin{equation}  \label{Eq.105}
dp^0=\left[ -\partial _i\left( \stackrel{\_}{A}^i-\partial ^iA_0\right)
-2a^2\partial ^k\partial ^k\left( \partial _i\stackrel{\_}{A}^i-\partial
_i\partial ^iA_0\right) \right] dt
\end{equation}

Finally for $\pi ^i$ we have:
\begin{equation}  \label{Eq.106}
d\pi ^i=-\frac{\partial H_0^{\prime }}{\partial \stackrel{\_}{A}_i}dt-\frac{%
\partial H_1^{\prime }}{\partial \stackrel{\_}{A}_i}d\stackrel{\_}{A}_0-%
\frac{\partial H_2^{\prime }}{\partial \stackrel{\_}{A}_i}dA_0=- \frac{%
\partial H_0^{\prime }}{\partial \stackrel{\_}{A}_i}dt=-\frac{\partial H_C}{%
\partial \stackrel{\_}{A}_i}dt
\end{equation}

\begin{eqnarray}
d\pi ^i&=&-\int d^3x\left[ p^j\frac{\partial \stackrel{\_}{A}_j}{\partial
\stackrel{\_}{A}_i}+\left( \stackrel{\_}{A}^j-\partial ^jA_0\right) \frac{%
\partial \left( \stackrel{\_}{A}_j-\partial _jA_0\right) }{\partial A_0}%
\right.  \label{Eq.107} \\
&&\left. -2a^2\left( \partial _j\stackrel{\_}{A}^j-\partial _j\partial
^jA_0\right) \frac{\partial \left( \partial _k\stackrel{\_}{A}^k-\partial
_k\partial ^kA_0\right) }{\partial A_0}\right] dt  \nonumber \\
\nonumber
\end{eqnarray}
\begin{equation}  \label{Eq.108}
d\pi ^i=\left[ -p^i-F^{0i}-2a^2\partial ^i\partial _kF^{0k}\right] dt
\end{equation}

Equations (\ref{Eq.102}), (\ref{Eq.105}) and (\ref{Eq.108}) are completely
equivalent to (\ref{Eq.92}), (\ref{Eq.93}) and (\ref{Eq.94}); consequently
equations (\ref{Eq.105}) and (\ref{Eq.108}) give us the secondary constraint
that isn't present in the total diffe\-ren\-tial equations.

\section{Conclusions}

We obtained a generalization of Hamilton-Jacobi formalism whose results
agree with those obtained using Dirac's formalism. In this formalism those
coordinates whose correspondent accelerations can't be solved in function of
the momenta are arbitrary variables of the theory. We obtained a set of
Hamilton-Jacobi partial differential equations in terms of these variables
and from this set we obtained the equations of motion of the system as the
total differential equations for the characteristics. These total
differential equations so obtained must satisfy integrability conditions and
for these conditions to be satisfied the nature of the constraints (first
class or second class) will play an essential role. The study of these
integrability conditions is in progress as well as the generalization of the
present formalism for Lagrangians of order higher than two.

\end{document}